\documentclass[aps,pre,noshowkeys,
nosuperscriptaddress,
longbibliography,
reprint, 
nofootinbib,
floatfix,
]{revtex4-2}

\usepackage[utf8]{inputenc}
\usepackage{graphicx}
\usepackage[colorlinks=true,hyperfootnotes=true,breaklinks=true,linkcolor=blue,citecolor=blue]{hyperref}

\begin{document}
\title{Status achieved in an organization -- rank dynamics}

\author{Maciej Wołoszyn}
\thanks{\includegraphics[width=10pt]{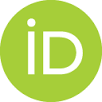}~\href{https://orcid.org/0000-0001-9896-1018}{0000-0001-9896-1018}}
\email{woloszyn@agh.edu.pl}
\affiliation{AGH University of Science and Technology,
Faculty of Physics and Applied Computer Science,
al. Mickiewicza 30, 30-059 Krak\'ow, Poland}

\author{Krzysztof Kułakowski}
\thanks{\includegraphics[width=10pt]{ORCID.png}~\href{https://orcid.org/0000-0003-1168-7883}{0000-0003-1168-7883}}
\email{kulakowski@fis.agh.edu.pl}
\affiliation{AGH University of Science and Technology,
Faculty of Physics and Applied Computer Science,
al. Mickiewicza 30, 30-059 Krak\'ow, Poland}

\date{\today}

\begin{abstract}
A new model of an evolution of ranks of employees due to staff turnover in an organization is designed.
If the rank is determined only by the performance, the rank shift of incumbents due to the turnover is proportional to the initial rank: the status of high staff is reduced only slightly. This effect that has been observed in the literature.
However, if the status also depends on some other variable, the positions of high staff may be greatly deteriorated.
In that case, the effect depends nonlinearly on the time $\tau$ between subsequent replacements of staff.
\end{abstract}
\maketitle

\section{Introduction\label{sec:intro}}

 Formation of a hierarchy of social status is an ubiquitous phenomenon in each social setting \cite{gru,clr,lin3,nanlin,kilduff}. As defined in the literature \cite{ebri}, the social status is 'the relative rank that an individual holds, with attendant rights, duties, and lifestyle, in a social hierarchy based upon honour or prestige'. This definition reveals an inter-subjective character of status, as it places sources of the social status in human minds. Further, scholars distinguish between ascribed (age, gender, race) and achieved marks of status (education, occupation, marital state etc.) \cite{ebri}. As here we are interested in status dynamics, our considerations are limited to the latter.\\
 
Our motivation for this work was initialized by the paper by Prato and Ferraro \cite{prafe}. There, two hypotheses have been formulated in the context of hiring new employees in organizations:\\
 H1.~Hiring high-status newcomers negatively affects incumbents' performance,\\
 H2.~High-status incumbents experience a lower performance decline than low-status incumbents\\
and both have been confirmed when analyzing data collected in 1996-2007 on hiring security analysts \cite{prafe}. Our first aim is to demonstrate that if the social status of individuals in an organization is based on their performance, these hypotheses appear in a natural way within the frames of rank modeling. The conceptual link from status to performance is justified in subsection \ref{subsec:performance} on the basis of literature. \\

However, if this link is taken without reservations, a set of parameters related to social status, such as performance, productivity, education, experience, centrality etc. \cite{cr,pocr}, is reduced to one ordinal variable. In subsection \ref{subsec:1Dnotenough} a piece of literature is cited to show that this setting is quite restrictive. To make the approach more realistic, in the next step the model is generalized to include a dynamical coupling of the social status also with another variable. In subsection \ref{subsec:productivity} we discuss productivity in this role. Other interpretations of the coupled variable will be mentioned in the Discussion (Section \ref{sec:discussion}). In subsection \ref{subsec:currentmodels}, theoretical approaches are listed which support the hypotheses H1 and H2. \\

Our aim here is to develop a model which is able to reproduce the hypotheses H1 and H2, and which can be generalized to take into account a contribution to the social status related to an additional variable. Namely, although the status is determined by performance, it is influenced also by productivity because performance and productivity are coupled by the model dynamics. In section \ref{sec:evolution}, the model is described in details. Without the coupling, the time evolution reduces to a discrete rank dynamics, given in subsection \ref{subsec:discrete}. The coupling itself is introduced in subsection \ref{subsec:continuous}. At this stage the model is fully analytical. In subsection \ref{subsec:combined}, the model is supplemented by adding a removal of an employee with the worst status and hiring a new employee. \\

Our main result is that the coupling between performance and productivity deteriorates also the position of high-status (low rank) incumbents. This new effect is shown in Section \ref{sec:results} to depend non-monotonously on the coupling constant. Further possible generalizations of the model are discussed in the last section.

\subsection{Status based on performance\label{subsec:performance}}

Social status has been shown to appear in laboratory experiments as a mediating variable between performance and job satisfaction \cite{luo} and greed \cite{greed}. Also, in \cite{klasa} a psychological factor of self-efficacy, internalized from the socioeconomic status of parents, has been found to influence the pupils' performance at school.
On the other hand, in \cite{jemm}, high or low status was randomly assigned to school children, and then the children took individually an anagram test; as a result, children with high status performed significantly better. Similarly in \cite{king}, direct connection between status (either measured or manipulated) and performance has been found in neurobiological experiments on police officers.  Yet we should add that a similar experimental setting, reported in \cite{tom}, has given a less clear result.\\

A theoretical link between social status and performance has been drawn in \cite{cec}. In accordance with its inter-subjective character, status leads to performance expectations. These expectations produce real differences in performance in a self-fulfilling way. The conclusion of this subsection is that a large contribution to the social status is produced by the performance on socially accepted aims. For a thorough discussion of experiments and theory and a vast list of references we refer to \cite{cec}. 

\subsection{One dimension is not enough\label{subsec:1Dnotenough}}

In \cite{phn}, statistical data on the period from 1978 to 2019 have been presented 
to show that less popular ministers in the Danish government were dismissed more often. 
However, even in this case the popularity cannot be treated as a single criterion, 
and the very existence of any single criterion is untenable.
In \cite{jac}, the list of characteristics of teachers in Chicago eligible for 
dismissal by their principals has been shown to include teacher absences, value 
added measures, gender and age; in particular, old males were laid off more often. 
Similarly in \cite{jenter}, extended data on more than three thousand CEO turnovers 
in 1993-2009 indicated that most dismissals have not been justified by those CEO's agency. 
In \cite{flepp}, an attempt has been made to distinguish between wise and unwise dismissal of a football coach, based on his actual work (wise) and the role of bad luck (unwise). 
Basing on data on European football, collected for five seasons (2013/14--2017/18), the authors have shown that unwise dismissals do not improve the performance on the pitch. Similar conclusions have been drawn from earlier data on basketball, where the main criterion of evaluation of a manager was winning, and not efficiency \cite{fizel}.
 Other factors which influence the probability of dismissal dependence 
on the performance of the person to be fired are: the horizon of investors \cite{goyal},
policy of unions \cite{han}, national culture \cite{urban},
and the type of board \cite{wie,graz}. In the latter case this dependence indicates 
a kind of collusion 
between CEOs and the board which hired them; to dismiss the same person is equivalent to an 
admission that the former decision was a fault, and therefore it is often evaded. 
On the other hand, hiring a new manager is known to rise costs of reconstruction of social 
contacts \cite{fizel,hoff}.\\

Our conclusion from this list of examples is twofold. First is that status -- however understood -- of candidates to dismiss plays an important role. On the other hand, it is hard to find a case when a decision of firing an individual is taken based on a single parameter.

\subsection{Status and productivity\label{subsec:productivity}}

Coupling of productivity and social status is well known in literature, both in micro- and macro-scale, and several problems have been discussed in this context. In \cite{gua}, cross-cultural data have been used to show a  correlation between productivity and social class. In \cite{fly}, productivity has been found to be optimal for flat hierarchy; the status differences have been then attained through frequent favor exchange.  In \cite{bre}, a coupling of social status with consumption has been shown to lead to over-productivity and a decline of human welfare. In \cite{par}, productivity of scientists in acquired companies has been shown to go down when they lose their status in the combined entity. Detrimental consequences of low status for productivity have also been reported in \cite{wei} and literature therein. These notional connections allow one to treat productivity as directly correlated with social status.  \\

\subsection{Current models of rank dynamics\label{subsec:currentmodels}}

In \cite{blumm}, the Langevin equation has been proposed for the normalized market shares of economic items. Besides the evolving shares, the items were endowed with fitnesses, i.e. time-independent intrinsic abilities to increase their market shares \cite{blumm}. In a steady state, the share of an item was a monotonic function of its fitness. Also, the shares of different items influenced each other only via the normalization of their sum to unity. A central result of \cite{blumm} was a classification of different systems according to stability of the related rankings, with the word usage in English and the citations of scientific papers as opposing and extreme examples. In any case, the volatility of shares of the top-ranked items was found to be less than that of the items with low shares \cite{blumm}.\\

The ranking dynamics of words in five European languages was also considered in \cite{cocho}. The time evolution of the ranks of words have been analyzed within a random walk model. There, low-ranked (the most frequent) words have been shown to vary only slightly, and the variation of words -- to increase with their rank. Similar model has been also applied to twelve data sets on sport rankings (\cite{mors} and references therein). Also there, in all cases the top players or teams have been found to change their ranks more slowly, than the others.\\

In \cite{ini}, the Authors analyzed data on the time dependence of 30 ranking lists of different origin. On this basis a classification of rankings has been designed, built on two parameters: $\it{(i)}$ the rank turnover $\sigma_t$, defined as the number of elements ever seen in the rank list until time $t$, normalized by the list length, and $\it{(ii)}$ the rank flux $F_t$, defined as the probability that an element enters or leaves the rank list at time $t$. When averaged over time, the mean rate of turnover and the flux appear to be correlated. Further, the ranks have been classified according to the mean turnover rate $\dot\sigma=\Delta\sigma/\Delta t$, from $\dot\sigma\sim 1$ (more open) to $\dot\sigma\sim 0$ (more closed) as extreme cases. The model designed in \cite{ini} includes two mechanisms, displacement of elements in rank positions and replacement of a randomly selected elements by new ones. It is characteristic that in most systems, top ranked elements maintain their ranks during the time evolution \cite{ini}.\\

Concluding this subsection, the hypotheses H1 and H2 find at least a qualitative confirmation by the model results of \cite{blumm,cocho,mors,ini}.

\section{Time evolution\label{sec:evolution}}

Below we will use some elements of \cite{ini}: the mechanism of replacements (with our additional condition that the element ranked as the worst is replaced), and the relation between the turnover and its rate; one is the time derivative of the other.\\ 

In this section we describe the model dynamics. In its simple version we assume that ranks of elements are given by the values of status, which is equivalent to performance. In this case the time evolution is reduced to the evolution of the rank itself. In the more extended version, the relation between rank and status is the same, but the status variable is additionally coupled with another observable, termed here as productivity.
The core of the model is that status and productivity enhance each other. \\

In subsection \ref{subsec:discrete} the simple version of the algorithm is described, where the scores on social status of individuals are presented in the form of a rank. The dynamics is driven by a replacement of an element with last (worst) rank by an element with rank randomly selected. We will show that this setting allows to reproduce the hypotheses H1 and H2 \cite{prafe}, listed in the Introduction. Two subsequent subsections \ref{subsec:continuous} and \ref{subsec:combined} are devoted to the full version of the algorithm which includes a sequence of two alternating stages. In \ref{subsec:continuous}, the continuous stage is described and solved analytically. The continuous evolution is periodically interrupted by the replacements, similarly to those described in \ref{subsec:discrete}; this stage is described in subsection \ref{subsec:combined}. Therefore the whole dynamics can be termed as 'piecewisely analytical'.

\subsection{Discrete rank dynamics\label{subsec:discrete}}

\begin{figure}
    \centering
    \includegraphics[width=\columnwidth]{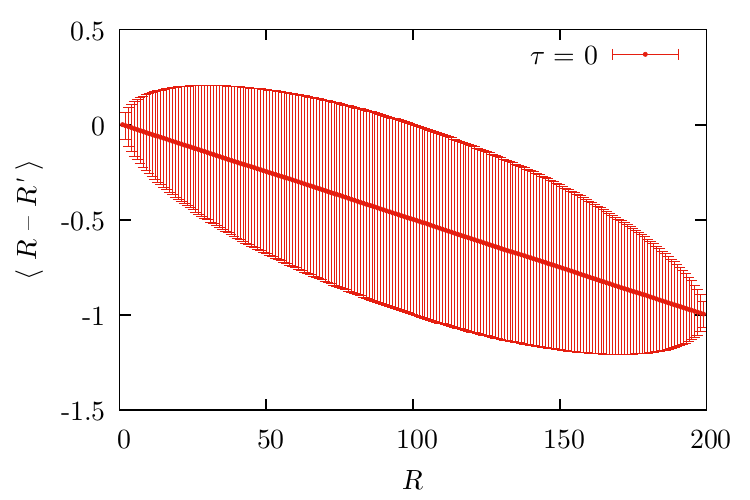}
    \caption{Mean shift of rank $\langle R - R' \rangle$ as dependent on rank $R$ for $\tau=0$, i.e. lack of the coupling. The vertical lines mark the standard deviations, added and subtracted. }
    \label{fig:tau0}
\end{figure}

In this version of the algorithm, there is no need to specify the scores which would be used to evaluate ranks. The number $N$ of elements remains constant. In each time step, the element at the last position $(R=N)$ in rank list is removed; instead, a new element is introduced with randomly selected rank $R'$ such that $R'<R$. As a consequence, each element of rank $R''$ such that $R'<R''$ is shifted down in the list, i.e. $R''\to R''+1$. The probability that an element of rank $R$ will be shifted is proportional to $R$. After $t_{max}$ time steps, the mean number of shifts is $Rt_{max}/N$.
This result is shown in Fig. \ref{fig:tau0}. Clearly, the status of incumbents decreases, and those at lower rank positions (i.e. of higher status) lose less, in accordance with the claims in literature \cite{prafe}.\\

In next subsections this replacement is described as entangled with the continuous evolution of performance and productivity. Both these variables are continuous. The rank list is formed on the basis of status, which is an inter-subjective measure of the performance.  

\subsection{Continuous dynamics\label{subsec:continuous}}
We consider a set of interacting actors. Each actor $i$ is endowed by a given productivity $y_i$. Also, a social status of actor $j$ in $i$'s mind is 
defined, $x_{ij}$. The essence of the model is in two causal assumptions on the dynamics:\\
-- the status $x_{ij}$ varies in time, proportionally to the difference of productivities,\\
-- the productivity $y_i$ of $i$ varies in time, proportionally to the $i$'th total status.\\
These assumptions are encoded in two model equations:
\begin{equation}
\frac{dy_{k}}{dt}=\alpha \sum_{j\neq k}x_{jk}
\end{equation}
\begin{equation}
\frac{dx_{jk}}{dt}=y_k-y_j
\end{equation}

In the Appendix \ref{app:A} we show that these equations are equivalent to

\begin{eqnarray}
\frac{dw_k}{dt}&=&v_k\\
\frac{dv_k}{dt}&=&w_k
\end{eqnarray}
where $w_k=y_k-\gamma$, $\gamma=(1/N) \sum_k y_k=const(t)$ and $\alpha$ is chosen to be $1/N$. The analytical solution is 
\begin{eqnarray}
w_k(t)&=&w_k(0) \cosh(t)+v_k(0) \sinh(t)\\
v_k(t)&=&w_k(0) \sinh(t)+v_k(0) \cosh(t)
\label{an}
\end{eqnarray}
In the matrix form it reads
\begin{equation}
\left( \begin{array}{c} w_k(t) \\ v_k(t) \end{array} \right)=A(t)\left( \begin{array}{c} w_k(0) \\ v_k(0) \end{array}\right)
\end{equation}
with the property $A(t_1+t_2)=A(t_1)A(t_2)$.\\

In this new representation, the variables $w_k$ are related to the productivities, and the variables $v_k$ to the statuses, based on the performances.\\

The transformation given in Eqn. (7), when applied several times, leads to a strong correlation of $w_k(t)$ and $v_k(t)$. This property is shown in the Appendix \ref{app:B}. The related number of steps depends on $t$; it is large for $t$ close to zero, and for large $t$ it decreases with $t$.

\subsection{Combined dynamics\label{subsec:combined}}

The continuous process described in the preceding subsection is assumed to be interrupted in discrete time moments $\tau,2\tau,3\tau,...$ etc. At each instant, an individual with minimal status is removed, and a new actor appears. The status of this actor is randomly chosen, and her/his productivity is inherited from the  individual just removed.\\

As we expect from Eqns. (5,6), the solution grows exponentially with time. Although the time scale of the process depends on particular application, the exponential growth can be numerically unfeasible. On the other hand, our aim is the dynamics of the related ranking, and the ranking order does not depend on the time scale. Then, 
the time evolution of $w_k$ and $v_k$ from $t=n\tau$ to $t=(n+1)\tau$ is supplemented by rescaling both sets of variables to the range $(-1,1)$. Namely,
\begin{eqnarray}
 v_k & \to & \frac{2v_k-v_M-v_m}{v_M-v_m} \\
 w_k & \to & \frac{2w_k-w_M-w_m}{w_M-w_m}
\end{eqnarray}
where $v_M$ and $w_M$ ($v_m$ and $w_m$) are the maximal (minimal) values of $v_k$ and $w_k$.
This mapping projects $v_m$ to $-1$, and $v_M$ to $+1$, and similarly $w_m$ to $-1$, and $w_M$ to $+1$. It preserves the homogeneous character of the distribution of $v$ values, $\rho(v)$.
Finally, a new value of $v_m$ is randomly chosen from the homogeneous distribution $\rho(v)=1/2$ for $-1<v<+1$, zero elsewhere. 
 This is done after each time step $\tau$. \\
 
 The detailed sequence of calculation steps is as follows:
 \begin{enumerate}
 \item select randomly $N$ values $v_k$ and $N$ values $w_k$ from the homogeneous distribution in the range $(-1,+1)$
 \item rescale both sets $v_k$ and $w_k$
 \item assign ranks $R(w_k)$ to the set $w_k$
 \item select randomly a number $g$ from the same distribution
 \item substitute $v_m$ by $g$ and rescale the set $v_k$ again
\item apply the transformation, based on Eqns. (5,6)
 \begin{eqnarray}
w_k(t+\tau)&=&w_k(t) \cosh(\tau)+v_k(t) \sinh(\tau)\\
v_k(t+\tau)&=&w_k(t) \sinh(\tau)+v_k(t) \cosh(\tau)
\end{eqnarray}
\item rescale the sets $v_k$ and $w_k$
\item assign ranks $R'(v_k)$ to the set $v_k$
\item compare $R$ and $R'$, as described in subsection \ref{subsec:discrete} (The ranking dynamics). Rename $R(v_k)=R'(v_k)$ for each $k$
\item go to 4
 \end{enumerate}
 The loop from 4 to 10 is repeated a prescribed number $t_{max}$ of time steps. \\

  \begin{figure}
    \centering
    \includegraphics[width=\columnwidth]{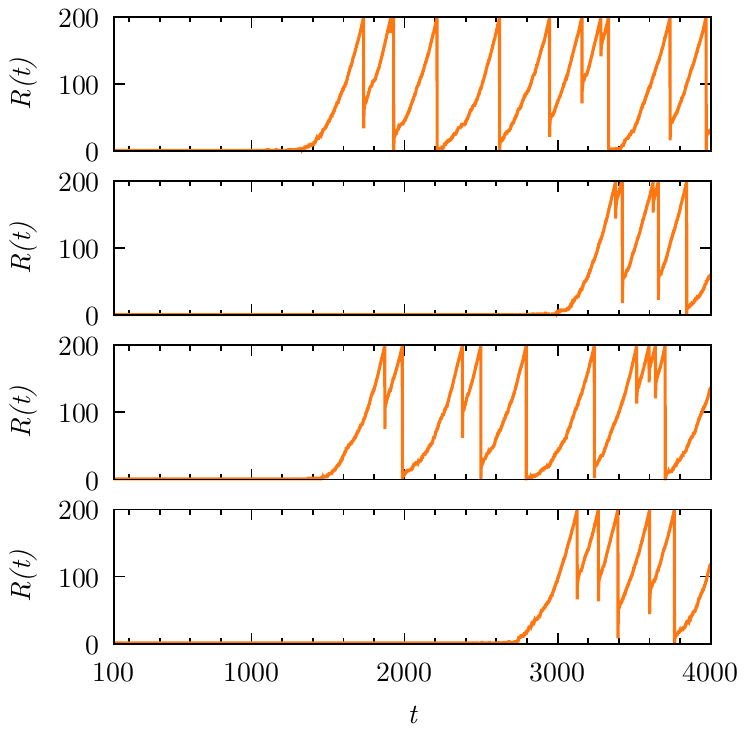}
    \caption{Examples of time evolution of a rank position of an agent, which started at $R=2$. Each time when the curve reaches $R=N$, the agent of this rank is removed and a new agent is introduced. Here $\tau=0.5$.}
    \label{ff0}
\end{figure}
 
 \begin{figure}
    \centering
    \includegraphics[width=\columnwidth]{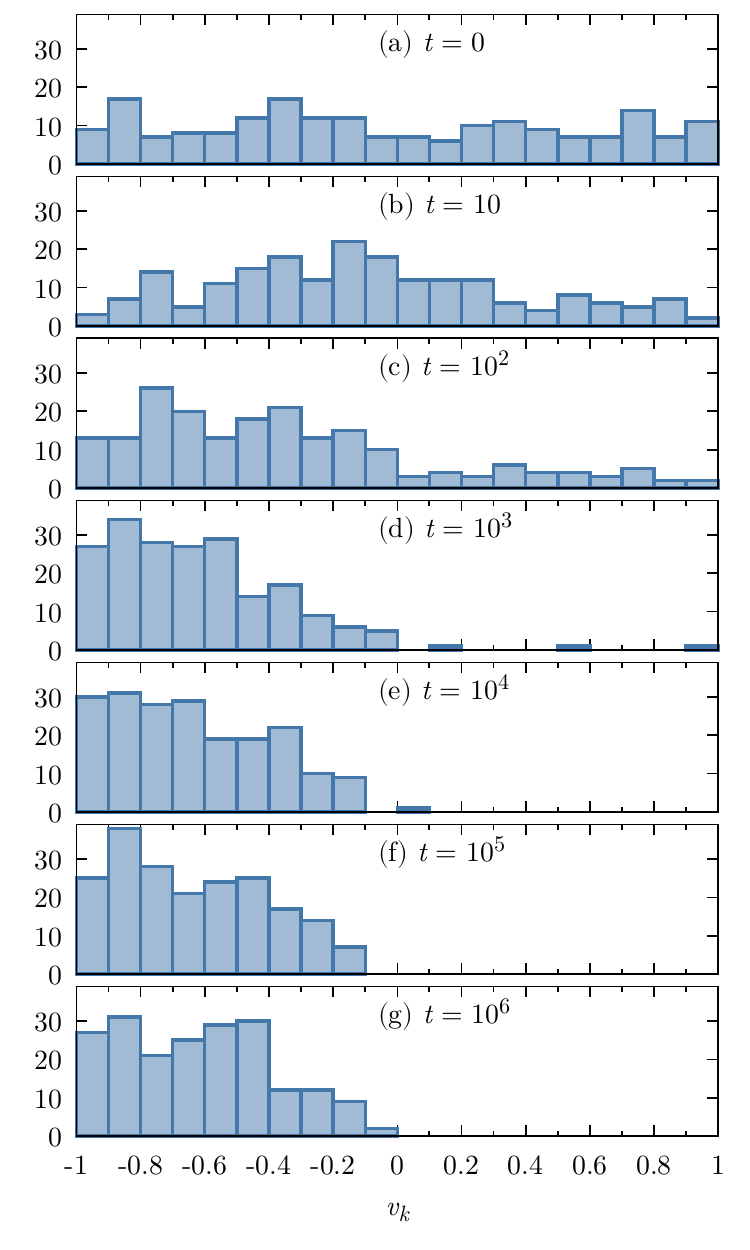}
    \caption{Time evolution of the distribution of statuses $v_k$ for $\tau=0.5$. The same effect is observed for larger $\tau$. For $\tau=0$ the distribution remains unchanged in time. }
    \label{fig:X}
\end{figure}
 
\section{Results\label{sec:results}}

In analogy to the Lagrangian and Eulerian representations of liquids \cite{benn}, the time evolution of rankings can be observed in two ways. The more intuitive (Lagrangian) method is to ask about the time evolution of the rank of a given entity $k$. Another (Eulerian) way is to ask what is the rank change of an entity of a given rank $R$. Below we address both these ways.\\

Observations of the rank of a given entity, if sufficiently long, give results which are easily predictable. Sooner or later, the status of each entity (except the best one) will appear to be the worst, and this status will be substituted by a random variable. Examples of such sequences are shown in Fig.~\ref{ff0}. Such observations are related to a particular agent if the time duration measured by the number of time steps $t_{max}$ is not much longer than the number $N$ of actors. In this sense, the results on the dynamics of the rank of a given agent are related to this agent only when limited to a transient stage of the process. (For completeness, this condition is not preserved in Fig.~\ref{ff0}.) On the contrary, there is no time limit for observations of the rank changes of entities of a given rank. Below, we show the results of these rank changes.\\

 It is worthwhile to add a comment on some specific consequences of the adopted rule of evolution. It is always the individual with the worst status (the maximal $R$) who is eliminated. As a rule, both her performance $v_m$ and her productivity $w_m$ are low, as for $\tau>0$ these quantities attract each other to form her status. The newly hired individual gets a new performance, but inherits low productivity, as acquired in the evolution. Consequently, the distribution of statuses is shifted towards negative values. This effect is shown in Fig.~\ref{fig:X}. The rule "remove the last" (of the worst performance), natural from an economic perspective \cite{stee}, mirrors the Red Queen effect, adopted from a fable to ecology: "we must run as fast as we can, just to stay in place" \cite{carr,vv}. In the theory of organizations, modifications of risk tolerance \cite{bec} can be seen as a counterpart of this effect.\\

In Fig.~\ref{fig:tau0} the line inclined from zero to $-1$ is the mean value of the shift of rank $\langle R - R' \rangle$ vs $R$, obtained for lack of coupling, i.e. for $\tau=0$. This is precisely the case described in subsection \ref{subsec:discrete}. The vertical lines mark the standard deviation, added and subtracted. As expected, these results agree with the binomial distribution, where the variance is equal to $(1-R/N)R/N$. The inclined line agrees with both hypotheses H1 and H2: \\
-- the obtained values are negative ($R'>R$), which means that incumbents lose;\\
-- the shift is smaller for incumbents of high-status (i.e. of low rank $R$), which means that they lose less.\\ 

\begin{figure}
    \centering
    \includegraphics[width=\columnwidth]{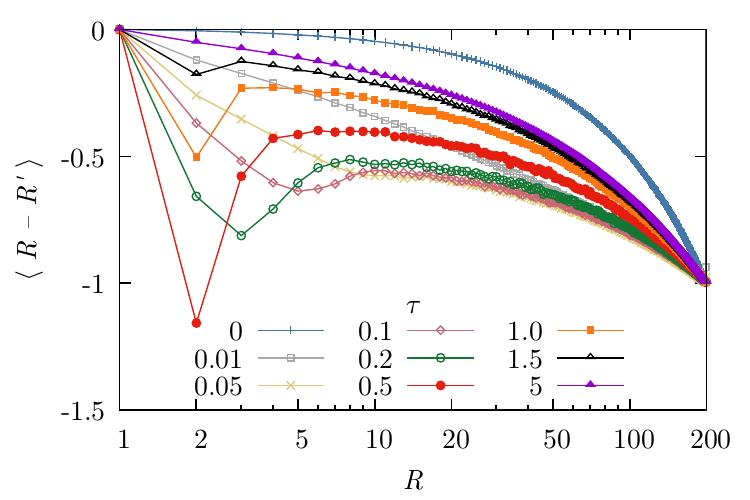}
    \caption{Mean shift $\langle R - R' \rangle$ as dependent on $R$ for various values of the coupling parameter $\tau$.}
    \label{coupling-mean}
\end{figure}

\begin{figure}
    \centering
    \includegraphics[width=\columnwidth]{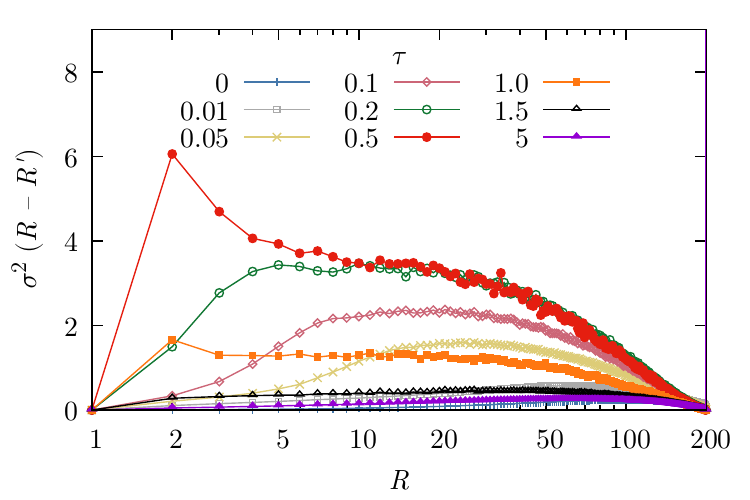}
    \caption{The variance of $\langle R - R' \rangle$ as dependent on $R$ for various values of the coupling parameter $\tau$.}
    \label{coupling-var}
\end{figure}

In the case of non-zero coupling ($\tau >0$), this picture is refined. The most important change appears for small values of $R$, where individuals of high status suffer an additional shift of rank. As shown in Fig.~\ref{coupling-mean}, the shift is no more linear with $R$. Instead, the plot shows an inflection point already for $\tau=0.05$. For higher $\tau$, there is a minimum of the plot $\langle R - R' \rangle$ vs $R$ (maximal shift) at some value of rank $R$ which decreases with $\tau$. However, for $\tau$ slightly above 1.5 the effect vanishes, and for $\tau =5.0$ the plot is similar to that of $\tau=0$. Similar irregularities are observed in the variance of the shift, as shown in Fig.~\ref{coupling-var}.\\

\begin{figure}
    \centering
    \includegraphics[width=\columnwidth]{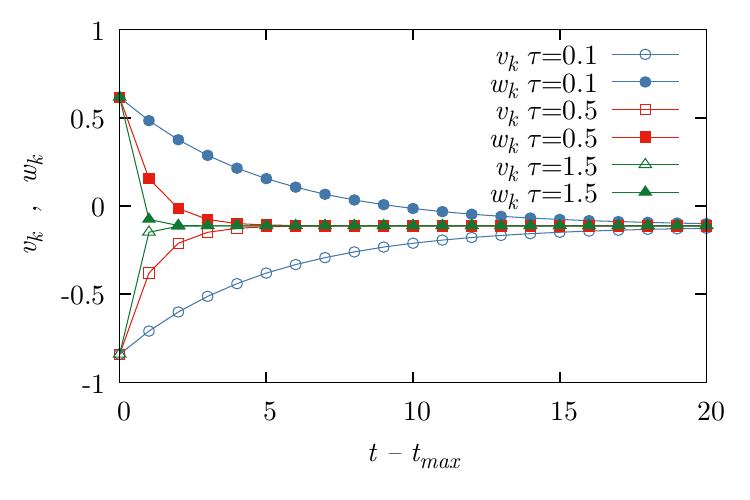}
    \caption{The time evolution of a pair $(w_k(t),v_k(t))$ according to the continuous dynamics, for various values of the coupling parameter $\tau$. The initial states are obtained with replacements of agents till $t=t_{max}$; later the replacements are abandoned. As we see, the time to level $v_k$ and $w_k$ decreases with $\tau$.}
    \label{y}
\end{figure}

The peculiarities of the mean shift of rank and its volatility, measured by its variance, in the range of low $R$, are a consequence of the time evolution of the distribution of the status $v_k$, shown in Fig.~\ref{fig:X}. Often, a newly introduced agent is endowed with a high status (low rank $R$). For small but non-zero $\tau$, the evolutions of the statuses of subsequently added agents overlap in time. This overlapping makes the evolution of low ranks complex. To visualize the characteristic time of evolution for different $\tau$, in Fig. \ref{y} we show an evolution of $v_k(t)$ and $v_k(t)$ without introducing new agents at each time step. 

\section{Discussion\label{sec:discussion}}

Within the simple version of the algorithm, described in subsection \ref{subsec:discrete}, we have identified the status with its measurable and dynamic aspect: performance. In this way, both hypotheses H1 and H2 of Prato and Ferraro \cite{prafe} have been confirmed. Within the expanded version of the algorithm, we have gained additional insight into the status dynamics. Namely, the coupling of status with an additional variable, termed above as productivity, has been shown to modify the rank dependence of the volatility of status. For some range of the coupling, controlled by the parameter $\tau$, the volatility is enhanced for individuals high in hierarchy (with low rank $R$).\\

In the literature, changes of employees' jobs within organization are termed internal mobility \cite{hass}. The distinction between internal mobility of high- and low-status employees is only rarely mentioned in the literature \cite{yy}. Internal mobility has been discussed as a way to reduce high staff turnover \cite{hass,hh} or as a marker of commitment \cite{lee}; the latter can be seen as correlated with high status \cite{stee}. Our result indicates that internal mobility is also a -- possibly unintended -- consequence of hiring new employees. Similar effects can be expected from internal training \cite{forr} and robotization \cite{figu}. \\ 

The role of the parameter $\tau$ is the intensity of the coupling of performance-based status with productivity. For $\tau=0$, there is no coupling and the results on status reduce to the simplest version of algorithm, as described in subsection \ref{subsec:discrete}. For large values of $\tau$, the coupling is strong, and status and productivity become equal after a small number of time steps. The volatility of individuals of high rank (small $R$) is most enhanced for small positive coupling $\tau$, as in this case the process of leveling status and productivity takes more time, and within this time subsequent reorderings of ranks disrupt each other. In any case, for $\tau>0$ the replacements of the worst elements give fuel to gradual equalization of the status and productivity of the new element. This is a difference when compared to Ref. \cite{ini}, where replacements and displacements were mutually independent mechanisms.\\

It is worth adding that our model formalism of status dynamics can be applied also to variables other than performance and productivity. The triad: status-power-access to resources is a well-established option \cite{clr,maxw}. Also there, elements of this triad mutually enhance each other. The difference is that in \cite{clr} status is treated mostly as an ascribed factor; as such, it contributes autonomously to causal relations with the other two. On the contrary, in our case an evaluation of status is based on measurable performance with its ups and downs, and therefore status ceases its stabilizing role described in \cite{clr,til}.  \\

According to this freedom of choice of model parameters, yet another possible application of our results can be drawn as follows. Suppose that formal statuses of employees in an organization, determined for example by their performance, are accompanied by an additional characteristics, as a position in an informal network of contacts; the latter not being legally recognized. As a rule, in a steady state the ranking of status takes both these characteristics into account, and both these characteristics of each worker fit to each other. Once a new employee appears, presumably of high performance, his position in the network is going to improve, which can lead to a partial reconstruction of the network. Accordingly, both formal and informal statuses of incumbents will be modified. Having recognized this process, organizations and employees attempt to control it by dedicated tools \cite{t1,t2,t3}.\\

We are not aware of a research which demonstrates how hiring new employees influences the rank structure of high staff. In the literature, the filling of jobs by external or internal candidates is usually presented as two alternative strategies of employers, sometimes termed 'buying versus making' \cite{bvm,bidk,ng}. Still, mutual interactions of consequences of these strategies are sometimes taken into account. In \cite{dlugos}, it has been shown that filling the job with an external (and not internal) candidate enhances the probability of exit of the rejected internal candidate. In \cite{hong}, the hiring of an external candidate was demonstrated to be profitable if accompanied by the reallocation of authority. This effect is already close to our topic of rank reconstruction. In \cite{raff,fros}, the authors refer to the so-called performance paradox. Its specific formulation 'external hires fail to replicate prior performance after switching firms' \cite{raff} reflects commonly known difficulties in prediction of performance. For scholars, its application to universities with quality identified with quantity \cite{fros} is of particular interest. Our results add to the unintended consequences of hiring external candidates, expressed in this formulation of the performance paradox.\\

A comment should be added on the algorithmic aspects of the model. As shown in the Appendix \ref{app:A}, between subsequent replacements the coupling between agents is removed and the calculation is fully analytical. The only stochastic process is related to the replacement of agents. This makes the model quite feasible numerically. It would be easy to generalize the approach, allowing for some distribution of times $\tau$ between the stochastic events.\\

Concluding, we have assumed that the social status of individuals in an organization is determined by their performance. This assumption allows reproducing the observations reported in the literature \cite{prafe,blumm,cocho,mors,ini}. Namely, newly hired individuals reduce the performance of incumbents, especially of those of low status. Furthermore, to take into account the multidimensional character of social status, the model is extended to include productivity, coupled with status with mutual positive feedback. This additional mechanism is shown to produce deterioration and high volatility of ranks in the range of high status. The effect can be seen as a mechanism of position mobility of high staff within an organization. A discussion of its cons and pros \cite{forr,kit,ma,col,fros} is beyond the scope of this paper.

\appendix

\section{\label{app:A}}

The equations (1,2) are equivalent to
\begin{equation}
 \frac{d^2y_k}{dt^2}=\alpha\sum_j(y_k-y_j)
\end{equation}

Next, suppose that at $t=0$ the sum of all statuses is zero:
\begin{equation}
 \sum_k\sum_{j\neq k} x_{jk}=0  
\end{equation}
Further, its time derivative is
\begin{equation}
\frac{d}{dt}\sum_k\sum_{j\neq k} x_{jk}
=\sum_k\sum_{j\neq k}(y_k-y_j)=(N-1)\sum_ky_k-\sum_k(\sum_jy_j-y_k)
\end{equation}
Having defined $\gamma=(1/N) \sum_ky_k$, we have
\begin{equation}
\frac{d}{dt}\sum_k\sum_{j\neq k} x_{jk}=(N-1)N\gamma-N^2\gamma+N\gamma=0
\end{equation}
Then we have $\sum_k\sum_{j\neq k} x_{jk}=0$ at each time. Therefore $d(\sum_ky_k)/dt=0$, i.e. $\gamma=const(t)$.\\

Furthermore, the two terms in r.h.s. of Eq. (3)
\begin{eqnarray}
 \sum_{j\ne k}y_k&=&(N-1)y_k \\
 \sum_{j\ne k}y_j&=&N\gamma-y_k
\end{eqnarray}
and, then,
\begin{equation}
 \sum_{j\ne k}(y_k-y_j)=Ny_k-y_k-N\gamma+y_k=N(y_k-\gamma)   
\end{equation}
We get
\begin{equation}
 \frac{d^2y_k}{dt^2}=\alpha N(y_k-\gamma)
\end{equation}
It is convenient to keep $\alpha N=1$, which defines the timescale of the whole process. With this choice, the time derivative of $y_k$ in Eq. (1) is roughly proportional to the mean status of the actor $k$. (In general, the dependence of the factor $\alpha$ on the system size $N$ depends on a particular application.) With new variables $w_k=y_k-\gamma$,
$v_k=dw_k/dt$ the basic equation of motion is 
\begin{equation}
 \frac{d^2w_k}{dt^2}=w_k  
\end{equation}
or, equivalently
\begin{eqnarray}
\frac{dw_k}{dt}&=&v_k\\
\frac{dv_k}{dt}&=&w_k
\end{eqnarray}
The analytical solution is 
\begin{eqnarray}
w_k(t)&=&w_k(0) \cosh(t)+v_k(0) \sinh(t)\\
v_k(t)&=&w_k(0) \sinh(t)+v_k(0) \cosh(t)
\label{ann}
\end{eqnarray}

\section{\label{app:B}}

Let us consider a pair of random variables $x,y$, both of mean value zero and the variances $\sigma^2_x$ and $\sigma^2_y$ equal 1, with the Pearson correlation coefficient between $x$ and $y$ equal to $f$.  Let us transform them to $x',y'$ such that
\begin{eqnarray}
x' & = & ax+by \nonumber\\
y' & = & bx+ay \nonumber
\end{eqnarray}
From a simple algebra, the variances of $x'$ and $y'$ are 
\begin{equation}
 \sigma^2_{x'} = \sigma^2_{y'} =a^2+b^2+2abf  
\end{equation}
and the correlation coefficient between the normalized $x'/\sigma_{x'}$ and $y'/\sigma_{y'}$ is 
\begin{equation}
 F=  \frac{2ab+(a^2+b^2)f}{a^2+b^2+2abf} 
\end{equation}
Having applied the transformation several times, we expect that $F=f$, which leads to $f^2=1$.\\
In our case $a= \cosh(\tau)$ and $b= \sinh(\tau)$. For $\tau=0$, the transformation is reduced to  identity. For large values of $\tau$, $x$ and $y$ are quickly mixed with each other.


\begin{thebibliography}{62}%
\makeatletter
\providecommand \@ifxundefined [1]{%
 \@ifx{#1\undefined}
}%
\providecommand \@ifnum [1]{%
 \ifnum #1\expandafter \@firstoftwo
 \else \expandafter \@secondoftwo
 \fi
}%
\providecommand \@ifx [1]{%
 \ifx #1\expandafter \@firstoftwo
 \else \expandafter \@secondoftwo
 \fi
}%
\providecommand \natexlab [1]{#1}%
\providecommand \enquote  [1]{``#1''}%
\providecommand \bibnamefont  [1]{#1}%
\providecommand \bibfnamefont [1]{#1}%
\providecommand \citenamefont [1]{#1}%
\providecommand \href@noop [0]{\@secondoftwo}%
\providecommand \href [0]{\begingroup \@sanitize@url \@href}%
\providecommand \@href[1]{\@@startlink{#1}\@@href}%
\providecommand \@@href[1]{\endgroup#1\@@endlink}%
\providecommand \@sanitize@url [0]{\catcode `\\12\catcode `\$12\catcode
  `\&12\catcode `\#12\catcode `\^12\catcode `\_12\catcode `\%12\relax}%
\providecommand \@@startlink[1]{}%
\providecommand \@@endlink[0]{}%
\providecommand \url  [0]{\begingroup\@sanitize@url \@url }%
\providecommand \@url [1]{\endgroup\@href {#1}{\urlprefix }}%
\providecommand \urlprefix  [0]{URL }%
\providecommand \Eprint [0]{\href }%
\providecommand \doibase [0]{https://doi.org/}%
\providecommand \selectlanguage [0]{\@gobble}%
\providecommand \bibinfo  [0]{\@secondoftwo}%
\providecommand \bibfield  [0]{\@secondoftwo}%
\providecommand \translation [1]{[#1]}%
\providecommand \BibitemOpen [0]{}%
\providecommand \bibitemStop [0]{}%
\providecommand \bibitemNoStop [0]{.\EOS\space}%
\providecommand \EOS [0]{\spacefactor3000\relax}%
\providecommand \BibitemShut  [1]{\csname bibitem#1\endcsname}%
\let\auto@bib@innerbib\@empty
\bibitem [{\citenamefont {Blau}\ and\ \citenamefont {Duncan}(2001)}]{gru}%
  \BibitemOpen
  \bibfield  {author} {\bibinfo {author} {\bibfnamefont {P.~M.}\ \bibnamefont
  {Blau}}\ and\ \bibinfo {author} {\bibfnamefont {O.~D.}\ \bibnamefont
  {Duncan}},\ }\bibfield  {title} {\bibinfo {title} {The process of
  stratification},\ }in\ \href {https://doi.org/10.4324/9780429494642} {\emph
  {\bibinfo {booktitle} {Social Stratification: Class, Race, and Gender in
  Sociological Perspective}}},\ \bibinfo {editor} {edited by\ \bibinfo {editor}
  {\bibfnamefont {D.~B.}\ \bibnamefont {Grusky}}}\ (\bibinfo  {publisher}
  {Westview Press},\ \bibinfo {address} {Boulder, Oxford},\ \bibinfo {year}
  {2001})\ p.\ \bibinfo {pages} {390}\BibitemShut {NoStop}%
\bibitem [{\citenamefont {Ridgeway}(2014)}]{clr}%
  \BibitemOpen
  \bibfield  {author} {\bibinfo {author} {\bibfnamefont {C.~L.}\ \bibnamefont
  {Ridgeway}},\ }\bibfield  {title} {\bibinfo {title} {Why status matters for
  inequality},\ }\href {https://doi.org/10.1177/0003122413515997} {\bibfield
  {journal} {\bibinfo  {journal} {Am. Sociol. Rev.}\ }\textbf {\bibinfo
  {volume} {79}},\ \bibinfo {pages} {1} (\bibinfo {year} {2014})}\BibitemShut
  {NoStop}%
\bibitem [{\citenamefont {Lin}\ \emph {et~al.}(1981)\citenamefont {Lin},
  \citenamefont {Vaughn},\ and\ \citenamefont {Ensel}}]{lin3}%
  \BibitemOpen
  \bibfield  {author} {\bibinfo {author} {\bibfnamefont {N.}~\bibnamefont
  {Lin}}, \bibinfo {author} {\bibfnamefont {J.~C.}\ \bibnamefont {Vaughn}},\
  and\ \bibinfo {author} {\bibfnamefont {W.~M.}\ \bibnamefont {Ensel}},\
  }\bibfield  {title} {\bibinfo {title} {Social resources and occupational
  status attainment},\ }\href {https://doi.org/10.2307/2577987} {\bibfield
  {journal} {\bibinfo  {journal} {Soc. Forces}\ }\textbf {\bibinfo {volume}
  {59}},\ \bibinfo {pages} {1163} (\bibinfo {year} {1981})}\BibitemShut
  {NoStop}%
\bibitem [{\citenamefont {Lin}(1999)}]{nanlin}%
  \BibitemOpen
  \bibfield  {author} {\bibinfo {author} {\bibfnamefont {N.}~\bibnamefont
  {Lin}},\ }\bibfield  {title} {\bibinfo {title} {Social networks and status
  attainment},\ }\href {https://doi.org/10.1146/annurev.soc.25.1.467}
  {\bibfield  {journal} {\bibinfo  {journal} {Annu. Rev. Sociol.}\ }\textbf
  {\bibinfo {volume} {25}},\ \bibinfo {pages} {467} (\bibinfo {year}
  {1999})}\BibitemShut {NoStop}%
\bibitem [{\citenamefont {Kilduff}\ \emph {et~al.}(2016)\citenamefont
  {Kilduff}, \citenamefont {Willer},\ and\ \citenamefont {Anderson}}]{kilduff}%
  \BibitemOpen
  \bibfield  {author} {\bibinfo {author} {\bibfnamefont {G.~J.}\ \bibnamefont
  {Kilduff}}, \bibinfo {author} {\bibfnamefont {R.}~\bibnamefont {Willer}},\
  and\ \bibinfo {author} {\bibfnamefont {C.}~\bibnamefont {Anderson}},\
  }\bibfield  {title} {\bibinfo {title} {Hierarchy and its discontents: status
  disagreement leads to withdrawal of contribution and lower group
  performance},\ }\href {https://doi.org/10.1287/orsc.2016.1058} {\bibfield
  {journal} {\bibinfo  {journal} {Organ. Sci.}\ }\textbf {\bibinfo {volume}
  {27}},\ \bibinfo {pages} {373} (\bibinfo {year} {2016})}\BibitemShut
  {NoStop}%
\bibitem [{ebr()}]{ebri}%
  \BibitemOpen
  \href@noop {} {\bibinfo {title} {Social status}},\ \bibinfo {howpublished}
  {\url{www.britannica.com/topic/social-status}},\ \bibinfo {note} {accessed:
  2022-07-23}\BibitemShut {NoStop}%
\bibitem [{\citenamefont {Prato}\ and\ \citenamefont {Ferraro}(2018)}]{prafe}%
  \BibitemOpen
  \bibfield  {author} {\bibinfo {author} {\bibfnamefont {M.}~\bibnamefont
  {Prato}}\ and\ \bibinfo {author} {\bibfnamefont {F.}~\bibnamefont
  {Ferraro}},\ }\bibfield  {title} {\bibinfo {title} {Starstruck: How hiring
  high-status employees affects incumbents' performance},\ }\href
  {https://doi.org/10.1287/orsc.2018.1204} {\bibfield  {journal} {\bibinfo
  {journal} {Organ. Sci.}\ }\textbf {\bibinfo {volume} {29}},\ \bibinfo {pages}
  {755} (\bibinfo {year} {2018})}\BibitemShut {NoStop}%
\bibitem [{\citenamefont {Chen}\ \emph {et~al.}(2012)\citenamefont {Chen},
  \citenamefont {Peterson}, \citenamefont {Phillips}, \citenamefont {Podolny},\
  and\ \citenamefont {Ridgeway}}]{cr}%
  \BibitemOpen
  \bibfield  {author} {\bibinfo {author} {\bibfnamefont {Y.~R.}\ \bibnamefont
  {Chen}}, \bibinfo {author} {\bibfnamefont {R.~S.}\ \bibnamefont {Peterson}},
  \bibinfo {author} {\bibfnamefont {D.~J.}\ \bibnamefont {Phillips}}, \bibinfo
  {author} {\bibfnamefont {J.~M.}\ \bibnamefont {Podolny}},\ and\ \bibinfo
  {author} {\bibfnamefont {C.~L.}\ \bibnamefont {Ridgeway}},\ }\bibfield
  {title} {\bibinfo {title} {Introduction to the special issue: bringing status
  to the table-attaining, maintaining, and experiencing status in organizations
  and markets},\ }\href {https://doi.org/10.1287/orsc.1110.0668} {\bibfield
  {journal} {\bibinfo  {journal} {Organ. Sci.}\ }\textbf {\bibinfo {volume}
  {23}},\ \bibinfo {pages} {299} (\bibinfo {year} {2012})}\BibitemShut
  {NoStop}%
\bibitem [{\citenamefont {Hong}\ \emph {et~al.}(2019)\citenamefont {Hong},
  \citenamefont {Zhang}, \citenamefont {Gang},\ and\ \citenamefont
  {Choi}}]{pocr}%
  \BibitemOpen
  \bibfield  {author} {\bibinfo {author} {\bibfnamefont {W.}~\bibnamefont
  {Hong}}, \bibinfo {author} {\bibfnamefont {L.}~\bibnamefont {Zhang}},
  \bibinfo {author} {\bibfnamefont {K.}~\bibnamefont {Gang}},\ and\ \bibinfo
  {author} {\bibfnamefont {B.}~\bibnamefont {Choi}},\ }\bibfield  {title}
  {\bibinfo {title} {The effects of expertise and social status on team member
  influence and the moderating roles of intragroup conflicts},\ }\href
  {https://doi.org/10.1177/1059601117728145} {\bibfield  {journal} {\bibinfo
  {journal} {Group Organ. Manag.}\ }\textbf {\bibinfo {volume} {44}},\ \bibinfo
  {pages} {745} (\bibinfo {year} {2019})}\BibitemShut {NoStop}%
\bibitem [{\citenamefont {Luo}\ \emph {et~al.}(2019)\citenamefont {Luo},
  \citenamefont {Permzadian}, \citenamefont {Fan},\ and\ \citenamefont
  {Meng}}]{luo}%
  \BibitemOpen
  \bibfield  {author} {\bibinfo {author} {\bibfnamefont {Y.}~\bibnamefont
  {Luo}}, \bibinfo {author} {\bibfnamefont {V.}~\bibnamefont {Permzadian}},
  \bibinfo {author} {\bibfnamefont {J.~Y.}\ \bibnamefont {Fan}},\ and\ \bibinfo
  {author} {\bibfnamefont {H.}~\bibnamefont {Meng}},\ }\bibfield  {title}
  {\bibinfo {title} {Employees' social self-efficacy and work outcomes: testing
  the mediating role of social status},\ }\href
  {https://doi.org/10.1177/1069072718795401} {\bibfield  {journal} {\bibinfo
  {journal} {J. Career Assess.}\ }\textbf {\bibinfo {volume} {27}},\ \bibinfo
  {pages} {661} (\bibinfo {year} {2019})}\BibitemShut {NoStop}%
\bibitem [{\citenamefont {Zhu}\ \emph {et~al.}(2019)\citenamefont {Zhu},
  \citenamefont {Sun}, \citenamefont {Liu},\ and\ \citenamefont {Xue}}]{greed}%
  \BibitemOpen
  \bibfield  {author} {\bibinfo {author} {\bibfnamefont {Y.~M.}\ \bibnamefont
  {Zhu}}, \bibinfo {author} {\bibfnamefont {X.~M.}\ \bibnamefont {Sun}},
  \bibinfo {author} {\bibfnamefont {S.~J.}\ \bibnamefont {Liu}},\ and\ \bibinfo
  {author} {\bibfnamefont {G.}~\bibnamefont {Xue}},\ }\bibfield  {title}
  {\bibinfo {title} {Is greed a double-edged sword? {The} roles of the need for
  social status and perceived distributive justice in the relationship between
  greed and job performance},\ }\href
  {https://doi.org/10.3389/fpsyg.2019.02021} {\bibfield  {journal} {\bibinfo
  {journal} {Front. Psychol.}\ }\textbf {\bibinfo {volume} {10}},\ \bibinfo
  {pages} {2021} (\bibinfo {year} {2019})}\BibitemShut {NoStop}%
\bibitem [{\citenamefont {Wiederkehr}\ \emph {et~al.}(2015)\citenamefont
  {Wiederkehr}, \citenamefont {Darnon}, \citenamefont {Chazal}, \citenamefont
  {Guimond},\ and\ \citenamefont {Martinot}}]{klasa}%
  \BibitemOpen
  \bibfield  {author} {\bibinfo {author} {\bibfnamefont {V.}~\bibnamefont
  {Wiederkehr}}, \bibinfo {author} {\bibfnamefont {C.}~\bibnamefont {Darnon}},
  \bibinfo {author} {\bibfnamefont {S.}~\bibnamefont {Chazal}}, \bibinfo
  {author} {\bibfnamefont {S.}~\bibnamefont {Guimond}},\ and\ \bibinfo {author}
  {\bibfnamefont {D.}~\bibnamefont {Martinot}},\ }\bibfield  {title} {\bibinfo
  {title} {From social class to self-efficacy: internalization of low social
  status pupils' school performance},\ }\href
  {https://doi.org/10.1007/s11218-015-9308-8} {\bibfield  {journal} {\bibinfo
  {journal} {Soc. Psychol. Educ.}\ }\textbf {\bibinfo {volume} {18}},\ \bibinfo
  {pages} {769} (\bibinfo {year} {2015})}\BibitemShut {NoStop}%
\bibitem [{\citenamefont {Jemmott}\ and\ \citenamefont
  {Gonzalez}(1989)}]{jemm}%
  \BibitemOpen
  \bibfield  {author} {\bibinfo {author} {\bibfnamefont {J.~B.}\ \bibnamefont
  {Jemmott}}\ and\ \bibinfo {author} {\bibfnamefont {E.}~\bibnamefont
  {Gonzalez}},\ }\bibfield  {title} {\bibinfo {title} {Social-status, the
  status distribution, and performance in small-groups},\ }\href
  {https://doi.org/10.1111/j.1559-1816.1989.tb00271.x} {\bibfield  {journal}
  {\bibinfo  {journal} {J. Appl. Soc. Psychol.}\ }\textbf {\bibinfo {volume}
  {19}},\ \bibinfo {pages} {584} (\bibinfo {year} {1989})}\BibitemShut
  {NoStop}%
\bibitem [{\citenamefont {Akinola}\ and\ \citenamefont {Mendes}(2014)}]{king}%
  \BibitemOpen
  \bibfield  {author} {\bibinfo {author} {\bibfnamefont {M.}~\bibnamefont
  {Akinola}}\ and\ \bibinfo {author} {\bibfnamefont {W.~B.}\ \bibnamefont
  {Mendes}},\ }\bibfield  {title} {\bibinfo {title} {It's good to be the king:
  neurobiological benefits of higher social standing},\ }\href
  {https://doi.org/10.1177/1948550613485604} {\bibfield  {journal} {\bibinfo
  {journal} {Soc. Psychol. Pers. Sci.}\ }\textbf {\bibinfo {volume} {5}},\
  \bibinfo {pages} {43} (\bibinfo {year} {2014})}\BibitemShut {NoStop}%
\bibitem [{\citenamefont {Rutherford}(2004)}]{tom}%
  \BibitemOpen
  \bibfield  {author} {\bibinfo {author} {\bibfnamefont {M.~D.}\ \bibnamefont
  {Rutherford}},\ }\bibfield  {title} {\bibinfo {title} {The effect of social
  role on theory of mind reasoning},\ }\href
  {https://doi.org/10.1348/000712604322779488} {\bibfield  {journal} {\bibinfo
  {journal} {Br. J. Psychol.}\ }\textbf {\bibinfo {volume} {95}},\ \bibinfo
  {pages} {91} (\bibinfo {year} {2004})}\BibitemShut {NoStop}%
\bibitem [{\citenamefont {Ridgeway}(2001)}]{cec}%
  \BibitemOpen
  \bibfield  {author} {\bibinfo {author} {\bibfnamefont {C.~L.}\ \bibnamefont
  {Ridgeway}},\ }\bibfield  {title} {\bibinfo {title} {Social status and group
  structure},\ }in\ \href {https://doi.org/10.1002/9780470998458.ch15} {\emph
  {\bibinfo {booktitle} {Blackwell Handbook of Social Psychology: Group
  Processes}}},\ \bibinfo {editor} {edited by\ \bibinfo {editor} {\bibfnamefont
  {M.~A.}\ \bibnamefont {Hogg}}\ and\ \bibinfo {editor} {\bibfnamefont {R.~S.}\
  \bibnamefont {Tindale}}}\ (\bibinfo  {publisher} {Blackwell Publishers Ltd},\
  \bibinfo {address} {Malden},\ \bibinfo {year} {2001})\BibitemShut {NoStop}%
\bibitem [{\citenamefont {Nielsen}(2022)}]{phn}%
  \BibitemOpen
  \bibfield  {author} {\bibinfo {author} {\bibfnamefont {P.~H.}\ \bibnamefont
  {Nielsen}},\ }\bibfield  {title} {\bibinfo {title} {Public popularity as a
  part of the job description? {Dismissing} unpopular ministers},\ }\href
  {https://doi.org/10.1080/01402382.2020.1837577} {\bibfield  {journal}
  {\bibinfo  {journal} {West Eur. Politics}\ }\textbf {\bibinfo {volume}
  {45}},\ \bibinfo {pages} {360} (\bibinfo {year} {2022})}\BibitemShut
  {NoStop}%
\bibitem [{\citenamefont {Jacob}(2010)}]{jac}%
  \BibitemOpen
  \bibfield  {author} {\bibinfo {author} {\bibfnamefont {B.~A.}\ \bibnamefont
  {Jacob}},\ }\href {https://doi.org/10.3386/w15715} {\bibinfo {title} {Do
  principals fire the worst teachers?}},\ \bibinfo {howpublished} {Nat. Bureau
  of Economic Research, Working paper 15715} (\bibinfo {year}
  {2010})\BibitemShut {NoStop}%
\bibitem [{\citenamefont {Jenter}\ and\ \citenamefont {Kanaan}(2015)}]{jenter}%
  \BibitemOpen
  \bibfield  {author} {\bibinfo {author} {\bibfnamefont {D.}~\bibnamefont
  {Jenter}}\ and\ \bibinfo {author} {\bibfnamefont {F.}~\bibnamefont
  {Kanaan}},\ }\bibfield  {title} {\bibinfo {title} {{CEO} turnover and
  relative performance evaluation},\ }\href
  {https://doi.org/10.1111/jofi.12282} {\bibfield  {journal} {\bibinfo
  {journal} {J. Finance}\ }\textbf {\bibinfo {volume} {70}},\ \bibinfo {pages}
  {2155} (\bibinfo {year} {2015})}\BibitemShut {NoStop}%
\bibitem [{\citenamefont {Flepp}\ and\ \citenamefont {Franck}(2001)}]{flepp}%
  \BibitemOpen
  \bibfield  {author} {\bibinfo {author} {\bibfnamefont {R.}~\bibnamefont
  {Flepp}}\ and\ \bibinfo {author} {\bibfnamefont {E.}~\bibnamefont {Franck}},\
  }\bibfield  {title} {\bibinfo {title} {The performance effects of wise and
  unwise managerial dismissals},\ }\href {https://doi.org/10.1111/ecin.12924}
  {\bibfield  {journal} {\bibinfo  {journal} {Econ. Inq.}\ }\textbf {\bibinfo
  {volume} {59}},\ \bibinfo {pages} {186} (\bibinfo {year} {2001})}\BibitemShut
  {NoStop}%
\bibitem [{\citenamefont {Fizel}\ and\ \citenamefont {D’Itri}(1999)}]{fizel}%
  \BibitemOpen
  \bibfield  {author} {\bibinfo {author} {\bibfnamefont {J.~L.}\ \bibnamefont
  {Fizel}}\ and\ \bibinfo {author} {\bibfnamefont {M.~P.}\ \bibnamefont
  {D’Itri}},\ }\bibfield  {title} {\bibinfo {title} {Firing and hiring
  managers: does efficiency matter?},\ }\href
  {https://doi.org/10.1177/014920639902500405} {\bibfield  {journal} {\bibinfo
  {journal} {J. Manag.}\ }\textbf {\bibinfo {volume} {25}},\ \bibinfo {pages}
  {567} (\bibinfo {year} {1999})}\BibitemShut {NoStop}%
\bibitem [{\citenamefont {Goyal}\ and\ \citenamefont {Low}(2019)}]{goyal}%
  \BibitemOpen
  \bibfield  {author} {\bibinfo {author} {\bibfnamefont {V.~K.}\ \bibnamefont
  {Goyal}}\ and\ \bibinfo {author} {\bibfnamefont {A.}~\bibnamefont {Low}},\
  }\bibfield  {title} {\bibinfo {title} {Investor myopia and {CEO} turnover},\
  }\href {https://doi.org/10.1111/irfi.12198} {\bibfield  {journal} {\bibinfo
  {journal} {Int. Rev. Finance}\ }\textbf {\bibinfo {volume} {19}},\ \bibinfo
  {pages} {759} (\bibinfo {year} {2019})}\BibitemShut {NoStop}%
\bibitem [{\citenamefont {Han}(2020)}]{han}%
  \BibitemOpen
  \bibfield  {author} {\bibinfo {author} {\bibfnamefont {E.~S.}\ \bibnamefont
  {Han}},\ }\bibfield  {title} {\bibinfo {title} {The myth of unions'
  overprotection of bad teachers: evidence from the district-teacher matched
  data on teacher turnover},\ }\href {https://doi.org/10.1111/irel.12256}
  {\bibfield  {journal} {\bibinfo  {journal} {Ind. Relat.}\ }\textbf {\bibinfo
  {volume} {59}},\ \bibinfo {pages} {316} (\bibinfo {year} {2020})}\BibitemShut
  {NoStop}%
\bibitem [{\citenamefont {Urban}(2019)}]{urban}%
  \BibitemOpen
  \bibfield  {author} {\bibinfo {author} {\bibfnamefont {D.}~\bibnamefont
  {Urban}},\ }\bibfield  {title} {\bibinfo {title} {The effects of culture on
  {CEO} power: evidence from executive turnover},\ }\href
  {https://doi.org/10.1016/j.jbankfin.2019.05.003} {\bibfield  {journal}
  {\bibinfo  {journal} {J. Bank. Finance}\ }\textbf {\bibinfo {volume} {104}},\
  \bibinfo {pages} {50} (\bibinfo {year} {2019})}\BibitemShut {NoStop}%
\bibitem [{\citenamefont {Wiersema}(2002)}]{wie}%
  \BibitemOpen
  \bibfield  {author} {\bibinfo {author} {\bibfnamefont {M.}~\bibnamefont
  {Wiersema}},\ }\bibfield  {title} {\bibinfo {title} {Holes at the top: why
  {CEO} firings backfire},\ }\href@noop {} {\bibfield  {journal} {\bibinfo
  {journal} {Harv. Bus. Rev.}\ }\textbf {\bibinfo {volume} {80}},\ \bibinfo
  {pages} {70} (\bibinfo {year} {2002})}\BibitemShut {NoStop}%
\bibitem [{\citenamefont {Graziano}\ and\ \citenamefont
  {Luporini}(2003)}]{graz}%
  \BibitemOpen
  \bibfield  {author} {\bibinfo {author} {\bibfnamefont {C.}~\bibnamefont
  {Graziano}}\ and\ \bibinfo {author} {\bibfnamefont {A.}~\bibnamefont
  {Luporini}},\ }\bibfield  {title} {\bibinfo {title} {Board efficiency and
  internal corporate control mechanisms},\ }\href
  {https://doi.org/10.1111/j.1430-9134.2003.00495.x} {\bibfield  {journal}
  {\bibinfo  {journal} {J. Econ. Manag. Strategy}\ }\textbf {\bibinfo {volume}
  {12}},\ \bibinfo {pages} {495} (\bibinfo {year} {2003})}\BibitemShut
  {NoStop}%
\bibitem [{\citenamefont {H{\"o}ffler}\ and\ \citenamefont
  {Sliwka}(2003)}]{hoff}%
  \BibitemOpen
  \bibfield  {author} {\bibinfo {author} {\bibfnamefont {F.}~\bibnamefont
  {H{\"o}ffler}}\ and\ \bibinfo {author} {\bibfnamefont {D.}~\bibnamefont
  {Sliwka}},\ }\bibfield  {title} {\bibinfo {title} {Do new brooms sweep clean?
  {When} and why dismissing a manager increases the subordinates’
  performance},\ }\href {https://doi.org/10.1016/S0014-2921(02)00272-6}
  {\bibfield  {journal} {\bibinfo  {journal} {Eur. Econ. Rev.}\ }\textbf
  {\bibinfo {volume} {47}},\ \bibinfo {pages} {877} (\bibinfo {year}
  {2003})}\BibitemShut {NoStop}%
\bibitem [{\citenamefont {Levine}\ \emph {et~al.}(2019)\citenamefont {Levine},
  \citenamefont {Harrington},\ and\ \citenamefont {Uhlmann}}]{gua}%
  \BibitemOpen
  \bibfield  {author} {\bibinfo {author} {\bibfnamefont {B.~R.}\ \bibnamefont
  {Levine}}, \bibinfo {author} {\bibfnamefont {J.~R.}\ \bibnamefont
  {Harrington}},\ and\ \bibinfo {author} {\bibfnamefont {E.~L.}\ \bibnamefont
  {Uhlmann}},\ }\bibfield  {title} {\bibinfo {title} {Culture and work},\ }in\
  \href@noop {} {\emph {\bibinfo {booktitle} {Handbook of Cultural
  Psychology}}},\ \bibinfo {editor} {edited by\ \bibinfo {editor}
  {\bibfnamefont {D.}~\bibnamefont {Cohen}}\ and\ \bibinfo {editor}
  {\bibfnamefont {S.}~\bibnamefont {Kitayama}}}\ (\bibinfo  {publisher} {The
  Guilford Press},\ \bibinfo {year} {2019})\ p.\ \bibinfo {pages}
  {630}\BibitemShut {NoStop}%
\bibitem [{\citenamefont {Flynn}(2003)}]{fly}%
  \BibitemOpen
  \bibfield  {author} {\bibinfo {author} {\bibfnamefont {F.~J.}\ \bibnamefont
  {Flynn}},\ }\bibfield  {title} {\bibinfo {title} {How much should {I} give
  and how often? the effects of generosity and frequency of favor exchange on
  social status and productivity},\ }\href {https://doi.org/10.5465/30040648}
  {\bibfield  {journal} {\bibinfo  {journal} {Acad. Manag. J.}\ }\textbf
  {\bibinfo {volume} {46}},\ \bibinfo {pages} {539} (\bibinfo {year}
  {2003})}\BibitemShut {NoStop}%
\bibitem [{\citenamefont {Brekke}\ \emph {et~al.}(2003)\citenamefont {Brekke},
  \citenamefont {Howarth},\ and\ \citenamefont {Nyborg}}]{bre}%
  \BibitemOpen
  \bibfield  {author} {\bibinfo {author} {\bibfnamefont {K.~A.}\ \bibnamefont
  {Brekke}}, \bibinfo {author} {\bibfnamefont {R.~B.}\ \bibnamefont
  {Howarth}},\ and\ \bibinfo {author} {\bibfnamefont {K.}~\bibnamefont
  {Nyborg}},\ }\bibfield  {title} {\bibinfo {title} {Status-seeking and
  material affluence: evaluating the {Hirsch} hypothesis},\ }\href
  {https://doi.org/10.1016/S0921-8009(02)00262-8} {\bibfield  {journal}
  {\bibinfo  {journal} {Ecol. Econ.}\ }\textbf {\bibinfo {volume} {45}},\
  \bibinfo {pages} {29} (\bibinfo {year} {2003})}\BibitemShut {NoStop}%
\bibitem [{\citenamefont {Paruchuri}\ \emph {et~al.}(2006)\citenamefont
  {Paruchuri}, \citenamefont {Nerkar}, \citenamefont {C.},\ and\ \citenamefont
  {Hambrick}}]{par}%
  \BibitemOpen
  \bibfield  {author} {\bibinfo {author} {\bibfnamefont {S.}~\bibnamefont
  {Paruchuri}}, \bibinfo {author} {\bibfnamefont {A.}~\bibnamefont {Nerkar}},
  \bibinfo {author} {\bibfnamefont {D.}~\bibnamefont {C.}},\ and\ \bibinfo
  {author} {\bibnamefont {Hambrick}},\ }\bibfield  {title} {\bibinfo {title}
  {Acquisition integration and productivity losses in the technical core:
  Disruption of inventors in acquired companies},\ }\href
  {https://doi.org/10.1287/orsc.1060.0207} {\bibfield  {journal} {\bibinfo
  {journal} {Organ. Sci.}\ }\textbf {\bibinfo {volume} {17}},\ \bibinfo {pages}
  {545} (\bibinfo {year} {2006})}\BibitemShut {NoStop}%
\bibitem [{\citenamefont {Weiss}\ \emph {et~al.}(2022)\citenamefont {Weiss},
  \citenamefont {Greve},\ and\ \citenamefont {Kunzmann}}]{wei}%
  \BibitemOpen
  \bibfield  {author} {\bibinfo {author} {\bibfnamefont {D.}~\bibnamefont
  {Weiss}}, \bibinfo {author} {\bibfnamefont {W.}~\bibnamefont {Greve}},\ and\
  \bibinfo {author} {\bibfnamefont {U.}~\bibnamefont {Kunzmann}},\ }\bibfield
  {title} {\bibinfo {title} {Responses to social inequality across the life
  span: The role of social status and upward mobility beliefs},\ }\href
  {https://doi.org/10.1177/01650254221089615} {\bibfield  {journal} {\bibinfo
  {journal} {Int. J. Behav. Dev.}\ }\textbf {\bibinfo {volume} {46}},\ \bibinfo
  {pages} {261} (\bibinfo {year} {2022})}\BibitemShut {NoStop}%
\bibitem [{\citenamefont {Blumm}\ \emph {et~al.}(2012)\citenamefont {Blumm},
  \citenamefont {Ghoshal}, \citenamefont {Forro}, \citenamefont {Schich},
  \citenamefont {Bianconi}, \citenamefont {Bouchaud},\ and\ \citenamefont
  {Barabasi}}]{blumm}%
  \BibitemOpen
  \bibfield  {author} {\bibinfo {author} {\bibfnamefont {N.}~\bibnamefont
  {Blumm}}, \bibinfo {author} {\bibfnamefont {G.}~\bibnamefont {Ghoshal}},
  \bibinfo {author} {\bibfnamefont {Z.}~\bibnamefont {Forro}}, \bibinfo
  {author} {\bibfnamefont {M.}~\bibnamefont {Schich}}, \bibinfo {author}
  {\bibfnamefont {G.}~\bibnamefont {Bianconi}}, \bibinfo {author}
  {\bibfnamefont {J.-P.}\ \bibnamefont {Bouchaud}},\ and\ \bibinfo {author}
  {\bibfnamefont {A.-L.}\ \bibnamefont {Barabasi}},\ }\bibfield  {title}
  {\bibinfo {title} {Dynamics of ranking processes in complex systems},\ }\href
  {https://doi.org/10.1103/PhysRevLett.109.128701} {\bibfield  {journal}
  {\bibinfo  {journal} {Phys. Rev. Lett.}\ }\textbf {\bibinfo {volume} {109}},\
  \bibinfo {pages} {128701} (\bibinfo {year} {2012})}\BibitemShut {NoStop}%
\bibitem [{\citenamefont {Cocho}\ \emph {et~al.}(2015)\citenamefont {Cocho},
  \citenamefont {Flores}, \citenamefont {Gershenson}, \citenamefont {Pineda},\
  and\ \citenamefont {Sanchez}}]{cocho}%
  \BibitemOpen
  \bibfield  {author} {\bibinfo {author} {\bibfnamefont {G.}~\bibnamefont
  {Cocho}}, \bibinfo {author} {\bibfnamefont {J.}~\bibnamefont {Flores}},
  \bibinfo {author} {\bibfnamefont {C.}~\bibnamefont {Gershenson}}, \bibinfo
  {author} {\bibfnamefont {C.}~\bibnamefont {Pineda}},\ and\ \bibinfo {author}
  {\bibfnamefont {S.}~\bibnamefont {Sanchez}},\ }\bibfield  {title} {\bibinfo
  {title} {Rank diversity of languages: generic behavior in computational
  linguistics},\ }\href {https://doi.org/10.1371/journal.pone.0121898}
  {\bibfield  {journal} {\bibinfo  {journal} {PLoS ONE}\ }\textbf {\bibinfo
  {volume} {10}},\ \bibinfo {pages} {e0121898} (\bibinfo {year}
  {2015})}\BibitemShut {NoStop}%
\bibitem [{\citenamefont {Morales}\ \emph {et~al.}(2021)\citenamefont
  {Morales}, \citenamefont {Flores}, \citenamefont {Gershenson},\ and\
  \citenamefont {Pineda}}]{mors}%
  \BibitemOpen
  \bibfield  {author} {\bibinfo {author} {\bibfnamefont {J.}~\bibnamefont
  {Morales}}, \bibinfo {author} {\bibfnamefont {J.}~\bibnamefont {Flores}},
  \bibinfo {author} {\bibfnamefont {C.}~\bibnamefont {Gershenson}},\ and\
  \bibinfo {author} {\bibfnamefont {C.}~\bibnamefont {Pineda}},\ }\bibfield
  {title} {\bibinfo {title} {Statistical properties of rankings in sports and
  games},\ }\href {https://doi.org/10.1142/S0219525921500077} {\bibfield
  {journal} {\bibinfo  {journal} {Adv. Complex Syst.}\ }\textbf {\bibinfo
  {volume} {24}},\ \bibinfo {pages} {2150007} (\bibinfo {year}
  {2021})}\BibitemShut {NoStop}%
\bibitem [{\citenamefont {Iñiguez}\ \emph {et~al.}(2022)\citenamefont
  {Iñiguez}, \citenamefont {Pineda}, \citenamefont {Gershenson},\ and\
  \citenamefont {Barabási}}]{ini}%
  \BibitemOpen
  \bibfield  {author} {\bibinfo {author} {\bibfnamefont {G.}~\bibnamefont
  {Iñiguez}}, \bibinfo {author} {\bibfnamefont {C.}~\bibnamefont {Pineda}},
  \bibinfo {author} {\bibfnamefont {C.}~\bibnamefont {Gershenson}},\ and\
  \bibinfo {author} {\bibfnamefont {A.-L.}\ \bibnamefont {Barabási}},\
  }\bibfield  {title} {\bibinfo {title} {Dynamics of ranking},\ }\href
  {https://doi.org/10.1038/s41467-022-29256-x} {\bibfield  {journal} {\bibinfo
  {journal} {Nat. Commun.}\ }\textbf {\bibinfo {volume} {13}},\ \bibinfo
  {pages} {1646} (\bibinfo {year} {2022})}\BibitemShut {NoStop}%
\bibitem [{\citenamefont {Bennett}(2006)}]{benn}%
  \BibitemOpen
  \bibfield  {author} {\bibinfo {author} {\bibfnamefont {A.}~\bibnamefont
  {Bennett}},\ }\href@noop {} {\emph {\bibinfo {title} {Lagrangian Fluid
  Dynamics}}}\ (\bibinfo  {publisher} {Cambridge University Press},\ \bibinfo
  {address} {Cambridge},\ \bibinfo {year} {2006})\BibitemShut {NoStop}%
\bibitem [{\citenamefont {Steenbergen}\ and\ \citenamefont
  {Ellemers}(2009)}]{stee}%
  \BibitemOpen
  \bibfield  {author} {\bibinfo {author} {\bibfnamefont {E.~F.~V.}\
  \bibnamefont {Steenbergen}}\ and\ \bibinfo {author} {\bibfnamefont
  {N.}~\bibnamefont {Ellemers}},\ }\bibfield  {title} {\bibinfo {title}
  {Feeling committed to work: how specific forms of work-commitment predict
  work behavior and performance over time},\ }\href
  {https://doi.org/10.1080/08959280903248385} {\bibfield  {journal} {\bibinfo
  {journal} {Hum. Perform.}\ }\textbf {\bibinfo {volume} {22}},\ \bibinfo
  {pages} {410} (\bibinfo {year} {2009})}\BibitemShut {NoStop}%
\bibitem [{\citenamefont {Carroll}(1993)}]{carr}%
  \BibitemOpen
  \bibfield  {author} {\bibinfo {author} {\bibfnamefont {L.}~\bibnamefont
  {Carroll}},\ }\href@noop {} {\emph {\bibinfo {title} {Alice in Wonderland}}}\
  (\bibinfo  {publisher} {Wordsworth Editions},\ \bibinfo {address} {London},\
  \bibinfo {year} {1993})\BibitemShut {NoStop}%
\bibitem [{\citenamefont {Valen}(1973)}]{vv}%
  \BibitemOpen
  \bibfield  {author} {\bibinfo {author} {\bibfnamefont {L.~V.}\ \bibnamefont
  {Valen}},\ }\bibfield  {title} {\bibinfo {title} {A new evolutionary law},\
  }\href@noop {} {\bibfield  {journal} {\bibinfo  {journal} {Evol. Theory}\
  }\textbf {\bibinfo {volume} {1}},\ \bibinfo {pages} {1} (\bibinfo {year}
  {1973})}\BibitemShut {NoStop}%
\bibitem [{\citenamefont {Greve}(2008)}]{bec}%
  \BibitemOpen
  \bibfield  {author} {\bibinfo {author} {\bibfnamefont {H.~R.}\ \bibnamefont
  {Greve}},\ }\bibfield  {title} {\bibinfo {title} {Organizational routines and
  performance feedback},\ }in\ \href@noop {} {\emph {\bibinfo {booktitle}
  {Handbook of Organizational Routines}}},\ \bibinfo {editor} {edited by\
  \bibinfo {editor} {\bibfnamefont {M.~C.}\ \bibnamefont {Becker}}}\ (\bibinfo
  {publisher} {Edward Elgar Publishing},\ \bibinfo {address} {Cheltenham},\
  \bibinfo {year} {2008})\ p.\ \bibinfo {pages} {187}\BibitemShut {NoStop}%
\bibitem [{\citenamefont {Hassink}(1996)}]{hass}%
  \BibitemOpen
  \bibfield  {author} {\bibinfo {author} {\bibfnamefont {W.~H.~J.}\
  \bibnamefont {Hassink}},\ }\bibfield  {title} {\bibinfo {title} {An empirical
  note on job turnover and internal mobility of workers},\ }\href
  {https://doi.org/10.1016/0165-1765(96)00830-0} {\bibfield  {journal}
  {\bibinfo  {journal} {Econ. Lett.}\ }\textbf {\bibinfo {volume} {51}},\
  \bibinfo {pages} {339} (\bibinfo {year} {1996})}\BibitemShut {NoStop}%
\bibitem [{\citenamefont {Yan}\ and\ \citenamefont {Yue}(2009)}]{yy}%
  \BibitemOpen
  \bibfield  {author} {\bibinfo {author} {\bibfnamefont {H.}~\bibnamefont
  {Yan}}\ and\ \bibinfo {author} {\bibfnamefont {F.~F.}\ \bibnamefont {Yue}},\
  }\bibfield  {title} {\bibinfo {title} {Research of knowledge workers and
  traditional staff in human resources management based on internal mobility},\
  }in\ \href@noop {} {\emph {\bibinfo {booktitle} {Human Resources Management
  in the Knowledge Economy Era, Vols. I and II}}}\ (\bibinfo {year} {2009})\
  p.\ \bibinfo {pages} {1295}\BibitemShut {NoStop}%
\bibitem [{\citenamefont {Han}\ and\ \citenamefont {Hur}(2022)}]{hh}%
  \BibitemOpen
  \bibfield  {author} {\bibinfo {author} {\bibfnamefont {D.}~\bibnamefont
  {Han}}\ and\ \bibinfo {author} {\bibfnamefont {H.}~\bibnamefont {Hur}},\
  }\bibfield  {title} {\bibinfo {title} {Managing turnover of {STEM} teacher
  workforce},\ }\href {https://doi.org/10.1177/00131245211053562} {\bibfield
  {journal} {\bibinfo  {journal} {Educ. Urban Soc.}\ }\textbf {\bibinfo
  {volume} {54}},\ \bibinfo {pages} {205} (\bibinfo {year} {2022})}\BibitemShut
  {NoStop}%
\bibitem [{\citenamefont {Lee}\ \emph {et~al.}(2018)\citenamefont {Lee},
  \citenamefont {Sohn}, \citenamefont {Kim}, \citenamefont {Kwon},\ and\
  \citenamefont {Park}}]{lee}%
  \BibitemOpen
  \bibfield  {author} {\bibinfo {author} {\bibfnamefont {J.}~\bibnamefont
  {Lee}}, \bibinfo {author} {\bibfnamefont {Y.~W.}\ \bibnamefont {Sohn}},
  \bibinfo {author} {\bibfnamefont {M.}~\bibnamefont {Kim}}, \bibinfo {author}
  {\bibfnamefont {S.}~\bibnamefont {Kwon}},\ and\ \bibinfo {author}
  {\bibfnamefont {I.~J.}\ \bibnamefont {Park}},\ }\bibfield  {title} {\bibinfo
  {title} {Relative importance of human resource practices on affective
  commitment and turnover intention in {South Korea} and {United States}},\
  }\href {https://doi.org/10.3389/fpsyg.2018.00669} {\bibfield  {journal}
  {\bibinfo  {journal} {Front. Psychol.}\ }\textbf {\bibinfo {volume} {9}},\
  \bibinfo {pages} {669} (\bibinfo {year} {2018})}\BibitemShut {NoStop}%
\bibitem [{\citenamefont {Forrier}\ and\ \citenamefont {Sels}(2003)}]{forr}%
  \BibitemOpen
  \bibfield  {author} {\bibinfo {author} {\bibfnamefont {A.}~\bibnamefont
  {Forrier}}\ and\ \bibinfo {author} {\bibfnamefont {L.}~\bibnamefont {Sels}},\
  }\bibfield  {title} {\bibinfo {title} {Flexibility, turnover and training},\
  }\href {https://doi.org/10.1108/01437720310475402} {\bibfield  {journal}
  {\bibinfo  {journal} {Int. J. Manpow.}\ }\textbf {\bibinfo {volume} {24}},\
  \bibinfo {pages} {148} (\bibinfo {year} {2003})}\BibitemShut {NoStop}%
\bibitem [{\citenamefont {Figueiredo}\ and\ \citenamefont
  {Pinto}(2021)}]{figu}%
  \BibitemOpen
  \bibfield  {author} {\bibinfo {author} {\bibfnamefont {A.}~\bibnamefont
  {Figueiredo}}\ and\ \bibinfo {author} {\bibfnamefont {L.}~\bibnamefont
  {Pinto}},\ }\bibfield  {title} {\bibinfo {title} {Robotizing shared service
  centres: key challenges and outcomes},\ }\href
  {https://doi.org/10.1108/JSTP-06-2020-0126} {\bibfield  {journal} {\bibinfo
  {journal} {J. Serv. Theory Pract.}\ }\textbf {\bibinfo {volume} {31}},\
  \bibinfo {pages} {157} (\bibinfo {year} {2021})}\BibitemShut {NoStop}%
\bibitem [{\citenamefont {Weber}(1978)}]{maxw}%
  \BibitemOpen
  \bibfield  {author} {\bibinfo {author} {\bibfnamefont {M.}~\bibnamefont
  {Weber}},\ }\href@noop {} {\emph {\bibinfo {title} {Economy and Society: An
  Outline of Interpretive Sociology}}}\ (\bibinfo  {publisher} {University of
  California Press},\ \bibinfo {address} {Berkeley},\ \bibinfo {year}
  {1978})\BibitemShut {NoStop}%
\bibitem [{\citenamefont {Tilly}(1998)}]{til}%
  \BibitemOpen
  \bibfield  {author} {\bibinfo {author} {\bibfnamefont {C.}~\bibnamefont
  {Tilly}},\ }\href@noop {} {\emph {\bibinfo {title} {Durable Inequality}}}\
  (\bibinfo  {publisher} {University of California Press},\ \bibinfo {address}
  {Berkeley},\ \bibinfo {year} {1998})\BibitemShut {NoStop}%
\bibitem [{\citenamefont {Leidner}\ \emph {et~al.}(2018)\citenamefont
  {Leidner}, \citenamefont {Gonzalez},\ and\ \citenamefont {Koch}}]{t1}%
  \BibitemOpen
  \bibfield  {author} {\bibinfo {author} {\bibfnamefont {D.~E.}\ \bibnamefont
  {Leidner}}, \bibinfo {author} {\bibfnamefont {E.}~\bibnamefont {Gonzalez}},\
  and\ \bibinfo {author} {\bibfnamefont {H.}~\bibnamefont {Koch}},\ }\bibfield
  {title} {\bibinfo {title} {An affordance perspective of enterprise social
  media and organizational socialization},\ }\href
  {https://doi.org/10.1016/j.jsis.2018.03.003} {\bibfield  {journal} {\bibinfo
  {journal} {J. Strateg. Inf. Syst.}\ }\textbf {\bibinfo {volume} {27}},\
  \bibinfo {pages} {117} (\bibinfo {year} {2018})}\BibitemShut {NoStop}%
\bibitem [{\citenamefont {Bolisani}\ and\ \citenamefont {Scarso}(2017)}]{t2}%
  \BibitemOpen
  \bibfield  {author} {\bibinfo {author} {\bibfnamefont {E.}~\bibnamefont
  {Bolisani}}\ and\ \bibinfo {author} {\bibfnamefont {E.}~\bibnamefont
  {Scarso}},\ }\bibfield  {title} {\bibinfo {title} {Exploring the use of an
  enterprise social network as a knowledge management tool in a medium-sized
  enterprise},\ }in\ \href@noop {} {\emph {\bibinfo {booktitle} {Proc. of the
  14th Int. Conf. on Intellectual Capital, Knowledge Management and
  Organisational Learning (ICICKM 2017)}}}\ (\bibinfo {year} {2017})\
  p.~\bibinfo {pages} {10}\BibitemShut {NoStop}%
\bibitem [{\citenamefont {Duparc}(2012)}]{t3}%
  \BibitemOpen
  \bibfield  {author} {\bibinfo {author} {\bibfnamefont {D.}~\bibnamefont
  {Duparc}},\ }\bibfield  {title} {\bibinfo {title} {Information and
  communication technologies: their impact on management and key functions to
  adapt},\ }in\ \href@noop {} {\emph {\bibinfo {booktitle} {Proc. of The 8th
  European Conf. on Management Leadership and Governance}}}\ (\bibinfo {year}
  {2012})\ p.\ \bibinfo {pages} {141}\BibitemShut {NoStop}%
\bibitem [{\citenamefont {Cooke}\ \emph {et~al.}()\citenamefont {Cooke},
  \citenamefont {Chowhan}, \citenamefont {{Mac Donald}},\ and\ \citenamefont
  {Mann}}]{bvm}%
  \BibitemOpen
  \bibfield  {author} {\bibinfo {author} {\bibfnamefont {G.~B.}\ \bibnamefont
  {Cooke}}, \bibinfo {author} {\bibfnamefont {J.}~\bibnamefont {Chowhan}},
  \bibinfo {author} {\bibfnamefont {K.}~\bibnamefont {{Mac Donald}}},\ and\
  \bibinfo {author} {\bibfnamefont {S.}~\bibnamefont {Mann}},\ }\bibfield
  {title} {\bibinfo {title} {Talent management: four 'buying versus making'
  talent development approaches},\ }\bibfield  {journal} {\bibinfo  {journal}
  {Pers. Rev.}\ }\href {https://doi.org/10.1108/PR-08-2020-0621}
  {10.1108/PR-08-2020-0621},\ \bibinfo {note} {in press}\BibitemShut {NoStop}%
\bibitem [{\citenamefont {Bidwell}\ and\ \citenamefont {Keller}(2014)}]{bidk}%
  \BibitemOpen
  \bibfield  {author} {\bibinfo {author} {\bibfnamefont {M.}~\bibnamefont
  {Bidwell}}\ and\ \bibinfo {author} {\bibfnamefont {J.~R.}\ \bibnamefont
  {Keller}},\ }\bibfield  {title} {\bibinfo {title} {Within or without? {How}
  firms combine internal and external labor markets to fill jobs},\ }\href
  {https://doi.org/10.5465/amj.2012.0119} {\bibfield  {journal} {\bibinfo
  {journal} {Acad. Manag. J.}\ }\textbf {\bibinfo {volume} {57}},\ \bibinfo
  {pages} {1035} (\bibinfo {year} {2014})}\BibitemShut {NoStop}%
\bibitem [{\citenamefont {Ng}\ and\ \citenamefont {Sherman}()}]{ng}%
  \BibitemOpen
  \bibfield  {author} {\bibinfo {author} {\bibfnamefont {W.}~\bibnamefont
  {Ng}}\ and\ \bibinfo {author} {\bibfnamefont {E.}~\bibnamefont {Sherman}},\
  }\bibfield  {title} {\bibinfo {title} {In search of inspiration: external
  mobility and the emergence of technology intrapreneurs},\ }\bibfield
  {journal} {\bibinfo  {journal} {Organ. Sci.}\ }\href
  {https://doi.org/10.1287/orsc.2021.1530} {10.1287/orsc.2021.1530},\ \bibinfo
  {note} {in press}\BibitemShut {NoStop}%
\bibitem [{\citenamefont {Dlugos}\ and\ \citenamefont {Keller}(2021)}]{dlugos}%
  \BibitemOpen
  \bibfield  {author} {\bibinfo {author} {\bibfnamefont {K.}~\bibnamefont
  {Dlugos}}\ and\ \bibinfo {author} {\bibfnamefont {J.~R.}\ \bibnamefont
  {Keller}},\ }\bibfield  {title} {\bibinfo {title} {Turned down and taking
  off? {Rejection} and turnover in internal talent markets},\ }\href
  {https://doi.org/10.5465/amj.2018.1015} {\bibfield  {journal} {\bibinfo
  {journal} {Acad. Manag. J.}\ }\textbf {\bibinfo {volume} {64}},\ \bibinfo
  {pages} {63} (\bibinfo {year} {2021})}\BibitemShut {NoStop}%
\bibitem [{\citenamefont {Hong}(2020)}]{hong}%
  \BibitemOpen
  \bibfield  {author} {\bibinfo {author} {\bibfnamefont {B.}~\bibnamefont
  {Hong}},\ }\bibfield  {title} {\bibinfo {title} {Power to the outsiders:
  external hiring and decision authority allocation within organizations},\
  }\href {https://doi.org/10.1002/smj.3182} {\bibfield  {journal} {\bibinfo
  {journal} {Strateg. Manag. J.}\ }\textbf {\bibinfo {volume} {41}},\ \bibinfo
  {pages} {1628} (\bibinfo {year} {2020})}\BibitemShut {NoStop}%
\bibitem [{\citenamefont {Raffiee}\ and\ \citenamefont {Byun}(2020)}]{raff}%
  \BibitemOpen
  \bibfield  {author} {\bibinfo {author} {\bibfnamefont {J.}~\bibnamefont
  {Raffiee}}\ and\ \bibinfo {author} {\bibfnamefont {H.}~\bibnamefont {Byun}},\
  }\bibfield  {title} {\bibinfo {title} {Revisiting the portability of
  performance paradox: employee mobility and the utilization of human and
  social capital resources},\ }\href {https://doi.org/10.5465/amj.2017.0769}
  {\bibfield  {journal} {\bibinfo  {journal} {Acad. Manag. J.}\ }\textbf
  {\bibinfo {volume} {63}},\ \bibinfo {pages} {34} (\bibinfo {year}
  {2020})}\BibitemShut {NoStop}%
\bibitem [{\citenamefont {Frost}\ and\ \citenamefont {Brockmann}(2014)}]{fros}%
  \BibitemOpen
  \bibfield  {author} {\bibinfo {author} {\bibfnamefont {J.}~\bibnamefont
  {Frost}}\ and\ \bibinfo {author} {\bibfnamefont {J.}~\bibnamefont
  {Brockmann}},\ }\bibfield  {title} {\bibinfo {title} {When qualitative
  productivity is equated with quantitative productivity: scholars caught in a
  performance paradox},\ }\href {https://doi.org/10.1007/s11618-014-0572-8}
  {\bibfield  {journal} {\bibinfo  {journal} {Z. Erziehungswiss.}\ }\textbf
  {\bibinfo {volume} {17}},\ \bibinfo {pages} {25} (\bibinfo {year}
  {2014})}\BibitemShut {NoStop}%
\bibitem [{\citenamefont {Kitowski}\ \emph {et~al.}(2004)\citenamefont
  {Kitowski}, \citenamefont {Krawczyk}, \citenamefont {Majewska}, \citenamefont
  {Dziewierz}, \citenamefont {Slota}, \citenamefont {Lambert}, \citenamefont
  {Miles}, \citenamefont {Arenas}, \citenamefont {Hluchy}, \citenamefont
  {Balogh}, \citenamefont {Laclavik}, \citenamefont {Delaitre}, \citenamefont
  {Viano}, \citenamefont {Stringa},\ and\ \citenamefont {Ferrentino}}]{kit}%
  \BibitemOpen
  \bibfield  {author} {\bibinfo {author} {\bibfnamefont {J.}~\bibnamefont
  {Kitowski}}, \bibinfo {author} {\bibfnamefont {K.}~\bibnamefont {Krawczyk}},
  \bibinfo {author} {\bibfnamefont {M.}~\bibnamefont {Majewska}}, \bibinfo
  {author} {\bibfnamefont {M.}~\bibnamefont {Dziewierz}}, \bibinfo {author}
  {\bibfnamefont {R.}~\bibnamefont {Slota}}, \bibinfo {author} {\bibfnamefont
  {S.}~\bibnamefont {Lambert}}, \bibinfo {author} {\bibfnamefont
  {A.}~\bibnamefont {Miles}}, \bibinfo {author} {\bibfnamefont
  {A.}~\bibnamefont {Arenas}}, \bibinfo {author} {\bibfnamefont
  {L.}~\bibnamefont {Hluchy}}, \bibinfo {author} {\bibfnamefont
  {Z.}~\bibnamefont {Balogh}}, \bibinfo {author} {\bibfnamefont
  {M.}~\bibnamefont {Laclavik}}, \bibinfo {author} {\bibfnamefont
  {S.}~\bibnamefont {Delaitre}}, \bibinfo {author} {\bibfnamefont
  {G.}~\bibnamefont {Viano}}, \bibinfo {author} {\bibfnamefont
  {S.}~\bibnamefont {Stringa}},\ and\ \bibinfo {author} {\bibfnamefont
  {P.}~\bibnamefont {Ferrentino}},\ }\bibfield  {title} {\bibinfo {title}
  {Model of experience for public organisations with staff mobility},\ }\href
  {https://doi.org/10.1007/978-3-540-24683-1_10} {\bibfield  {journal}
  {\bibinfo  {journal} {Lect. Notes Comput. Sci.}\ }\textbf {\bibinfo {volume}
  {3025}},\ \bibinfo {pages} {75} (\bibinfo {year} {2004})}\BibitemShut
  {NoStop}%
\bibitem [{\citenamefont {Jerala}\ and\ \citenamefont {Maček}(2012)}]{ma}%
  \BibitemOpen
  \bibfield  {author} {\bibinfo {author} {\bibfnamefont {M.}~\bibnamefont
  {Jerala}}\ and\ \bibinfo {author} {\bibfnamefont {M.~A.}\ \bibnamefont
  {Maček}},\ }\bibfield  {title} {\bibinfo {title} {How to improve quality of
  mobility at our institution},\ }in\ \href@noop {} {\emph {\bibinfo
  {booktitle} {Proc. of 5th International Conference of Education, Research And
  Innovation (ICERI 2012)}}}\ (\bibinfo {year} {2012})\ p.\ \bibinfo {pages}
  {2296}\BibitemShut {NoStop}%
\bibitem [{\citenamefont {Çolak}\ and\ \citenamefont
  {Korkeamäki}(2021)}]{col}%
  \BibitemOpen
  \bibfield  {author} {\bibinfo {author} {\bibfnamefont {G.}~\bibnamefont
  {Çolak}}\ and\ \bibinfo {author} {\bibfnamefont {T.}~\bibnamefont
  {Korkeamäki}},\ }\bibfield  {title} {\bibinfo {title} {{CEO} mobility and
  corporate policy risk},\ }\href
  {https://doi.org/10.1016/j.jcorpfin.2021.102037} {\bibfield  {journal}
  {\bibinfo  {journal} {J. Corp. Finance}\ }\textbf {\bibinfo {volume} {69}},\
  \bibinfo {pages} {102037} (\bibinfo {year} {2021})}\BibitemShut {NoStop}%
\end{thebibliography}

%


\end{document}